\documentclass[prb,draft,amsfonts,amssymb,amsmath,showpacs]{revtex4}
\begin{document}

\title{Dynamical Dirac Mass Generation in the Supersymmetric Nambu$-$Jona-Lasinio Model with the Seesaw Mechanism of Neutrinos}
\author{Tadafumi Ohsaku}
\affiliation{Institut f\"{u}r Theoretische Physik, Universit\"{a}t zu K\"{o}ln, 50937 K\"{o}ln, Germany}
\email[Present address: Department of Physics, University of Texas at Austin,]{tadafumi@physics.utexas.edu}
\date{\today}

\newcommand{\bmx}{\mbox{\boldmath $x$}}
\newcommand{\bmp}{\mbox{\boldmath $p$}}
\newcommand{\bmk}{\mbox{\boldmath $k$}}
\newcommand{\kfey}{\ooalign{\hfil/\hfil\crcr$k$}}
\newcommand{\pfey}{\ooalign{\hfil/\hfil\crcr$p$}}
\newcommand{\partfey}{\ooalign{\hfil/\hfil\crcr$\partial$}}
\newcommand{\hfey}{\ooalign{\hfil/\hfil\crcr$h$}}

\begin{abstract}

The dynamical generation of Dirac mass in the supersymmetric Nambu$-$Jona-Lasinio ( SNJL ) model with the seesaw mechanism of neutrino is investigeted.
The right and left handed Majorana mass parameters are introduced into the SNJL model; we regard them as external model parameters.
The question on the origin of these Majorana masses are set aside, 
and we concentrate on the examination of the effect of the Majorana mass parameters on the dynamical generation of Dirac mass.
The effective potential of the model and the gap equation for the self-consistent determination of Dirac mass are derived and solved.    
We use both the four-dimensional covariant and three-dimensional non-covariant cutoff schemes for the regularizations of the effective potential. 
We find there are cases of the first and second order phase transitions with respect to variation of the coupling constant of the Nambu$-$Jona-Lasinio-type four-body interaction of the SNJL model. In the case of second-order phase transition, the dynamically generated Dirac mass $|\phi_{S}|$ can arbitrarily be small compared with the right-handed Majorana mass parameter $|M|$ and thus the seesaw condition $0<|\phi_{S}|\ll|M|$ can be satisfied by a fine tuning of the coupling constant, 
while at the first-order case it seems very difficult and/or "unnatural" to satisfy the condition. 
The numerical results do not depend on the difference of the cutoff schemes qualitatively.

\end{abstract}


\maketitle

Recent experimental observations of flavor oscillations confirmed that neutrinos have small masses
( the mass of tau neutrino, $< 18$ MeV ) compared with other particles inside their same generations.
This fact motivates us to consider the seesaw mechanism of neutrinos~[1-3] more seriously,
because the mechanism seems to provide the unique explanation on tiny masses of neutrinos.
The seesaw mechanism states that Dirac masses ( for example, $\sim$ 300 GeV ) of neutrinos will be suppressed by very heavy Majorana masses ( for example, $2\times 10^{16}$ GeV ).
The electroweak gauge theory of the standard model is the most important success of modern particle physics,
which gives very accurate agreements with various experimental results, 
and still it will give us the horizon of particle phenomenology.
In the electroweak symmetry breaking, the top condensation model~[4,5] obtains much attractions from us until now,
because it is one of possible candidates toward beyond the standard model.
The top condensation scenario uses a generalization of the Nambu$-$Jona-Lasinio ( NJL ) model~[6] to include the electroweak gauge symmetry, 
and it considers a fermion-antifermion condensate. 
Recently, Antusch et al. gave a theory of an NJL-type four-body contact interaction model with introducing a right-handed Majorana mass parameter~[7].
In that work, the authors derived a gap equation for the dynamical generation of a Dirac mass term of neutrino, 
and solve them numerically.
Moreover, they argued the standard model with right-handed neutrinos as a low-energy effective theory of their model under its condensation scale $\Lambda$,
and performed a renormalization group analysis of model parameters of the renormalizable effective theory.
It is an interesting problem for us to examine a supersymmetric version of a four-fermion model combined with a right-handed Majorana mass term for neutrino.
This is the motivation of this paper.
We employ the supersymmetric Nambu$-$Jona-Lasinio ( SNJL ) model as our model Lagrangian.
Historically, the SNJL model was introduced for the investigation of dynamical chiral symmetry breaking in supersymmetric quantum field theory~[8], 
and it is extended to construct a supersymmetric version of the top condensation model~[9].
The SNJL model was also used to investigate the SUSY BCS-type superconductivity~[10,11].

\vspace{5mm}

We introduce the following one-flavor SNJL model with including the Majorana mass parameters:
\begin{eqnarray}
{\cal L} &\equiv& {\cal L}_{SNJL} + {\cal L}_{M} + {\cal L}_{m},  \\
{\cal L}_{SNJL} &\equiv&  \Bigl[ (1-\Delta^{2}\theta^{2}\bar{\theta}^{2})(\Phi^{\dagger}_{+}\Phi_{+} + \Phi^{\dagger}_{-}\Phi_{-}) + G\Phi^{\dagger}_{+}\Phi^{\dagger}_{-}\Phi_{+}\Phi_{-} \Bigr]_{\theta\theta\bar{\theta}\bar{\theta}}, \\
{\cal L}_{M} &\equiv& M^{\dagger}\Bigl[\Phi_{+}\Phi_{+}\Bigr]_{\theta\theta} + M\Bigl[\Phi^{\dagger}_{+}\Phi^{\dagger}_{+}\Bigr]_{\bar{\theta}\bar{\theta}},  \\ 
{\cal L}_{m} &\equiv& m^{\dagger}\Bigl[\Phi_{-}\Phi_{-}\Bigr]_{\theta\theta} + m\Bigl[\Phi^{\dagger}_{-}\Phi^{\dagger}_{-}\Bigr]_{\bar{\theta}\bar{\theta}}.
\end{eqnarray}
Here, $\Phi_{+}$ and $\Phi_{-}$ are right and left handed chiral superfields of neutrino, respectively.
We take the convention of spinor algebra by the book of Wess and Bagger~[12].
$M$ and $m$ are right- and left-handed Majorana mass parameters ( in general, given as complex numbers ), respectively.
The term of the coupling constant $G$ in (2) will be used as a four-body contact interaction similar to the case of the ordinary Nambu$-$Jona-Lasinio model~[6].
In four-dimensional spacetime, $G$ has mass dimension $[{\rm mass}]^{-2}$ and the theory is non-renormalizable.
In this paper, we perform the four-dimensional covariant and three-dimensional non-covariant cutoff regularizations, 
and a choice of a numerical value of the cutoff will be regarded as an introduction of the limit of the applicability of the theory by our hands. 
The electroweak gauge symmetry is not considered in this model for the sake of simplicity, 
hence it is a drastically simplified model prepared for our examination on the dynamical generation of a Dirac mass under including the Majorana mass parameters.
In the seesaw mechanism of neutrino masses, a right-handed Majorana mass of neutrino should take a huge value compared with its Dirac mass.
The mechanism of the generation of Majorana mass might be given in the different framework from the dynamical generation of Dirac mass
( for example, by a vacuum expectation value of a Higgs field ), hence we regard $M$ ( and also $m$ ) as a model parameter.
Of course, a simultaneous consideration of the dynamical generations of both the Dirac and the Majorana mass terms can be possible, 
by using the similar framework of the supersymmetric theory of superconductivity~[10,11].
Here, we assume the generation of the Majorana mass terms happens outside, a possible underlying theory, of our model.
In this paper, we just focus on the examination on some effects of the existence of the Majorana mass parameters in the dynamical generation of a Dirac mass term.
Our attitude to the problem is similar to the work of Ref.~[7], and we can say that our theory is a SUSY version of Ref.~[7],
though we do not consider the electroweak symmetry breaking.
Usually, it is believed that the existence of the SUSY is favorable to solve the problem of gauge hierarchy.
Because our model is determined as a one-flavor model, we do not consider the Kobayashi-Maskawa-like mixing matrix in our theory.
By using this model, we derive the effective potential and the gap equation for a self-consistent Dirac mass.

\vspace{5mm}

By using the method of SUSY auxiliary fields of composites~[8-11], the model Lagrangian will be converted into the following form:
\begin{eqnarray}
{\cal L} &=& \Bigl[  
(1-\Delta^{2}\theta^{2}\bar{\theta}^{2})(\Phi^{\dagger}_{+}\Phi_{+} + \Phi^{\dagger}_{-}\Phi_{-}) + \frac{1}{G}H^{\dagger}H \Bigr]_{\theta\theta\bar{\theta}\bar{\theta}}    \nonumber \\
& & + \Bigl[ S\Bigl( \frac{H}{G}-\Phi_{+}\Phi_{-} \Bigr) + M^{\dagger}\Phi_{+}\Phi_{+} + m^{\dagger}\Phi_{-}\Phi_{-}  \Bigr]_{\theta\theta}  + \Bigl[ S^{\dagger}\Bigl( \frac{H^{\dagger}}{G}- \Phi^{\dagger}_{+}\Phi^{\dagger}_{-} \Bigr) + M\Phi^{\dagger}_{+}\Phi^{\dagger}_{+} + m\Phi^{\dagger}_{-}\Phi^{\dagger}_{-} \Bigr]_{\bar{\theta}\bar{\theta}}.
\end{eqnarray}
Here, $S$ and $S^{\dagger}$ are Lagrange multiplier multiplets to keep the following definition of the composite collective chiral and antichiral superfields
$H=G\Phi_{+}\Phi_{-}$, $H^{\dagger}=G\Phi^{\dagger}_{+}\Phi^{\dagger}_{-}$, respectively.
Due to the non-renormalization theorem of supersymmetry, the ordinary SNJL model with keeping ${\cal N}=1$ SUSY cannot break the chiral symmetry dynamically~[8,9].
We consider it is also the case in our model.
In order to generate a Dirac mass term in our theory, we have introduced a SUSY breaking mass $\Delta$ in (5). 
Expanding ${\cal L}$ in terms of component fields, eliminating the auxiliary fields of chiral multiplets through their Euler-Lagrange equations, 
and assuming that the collective fields are spacetime-independent,
one finds the Lagrangian in the following form:
\begin{eqnarray}
{\cal L} &=& -\frac{|\phi_{S}|^{2}}{G} -\partial_{\nu}\phi^{\dagger}_{+}\partial^{\nu}\phi_{+} -\partial_{\nu}\phi^{\dagger}_{-}\partial^{\nu}\phi_{-}  \nonumber \\
& & - ( |\phi_{S}|^{2} + 4|M|^{2} + \Delta^{2} )\phi^{\dagger}_{+}\phi_{+} -( |\phi_{S}|^{2} + 4|m|^{2} + \Delta^{2} )\phi^{\dagger}_{-}\phi_{-}   \nonumber \\
& & + ( 2M\phi_{S} + 2m^{\dagger}\phi^{\dagger}_{S} )\phi^{\dagger}_{+}\phi_{-} + ( 2M^{\dagger}\phi^{\dagger}_{S} + 2m\phi_{S} )\phi^{\dagger}_{-}\phi_{+}  \nonumber \\
& & -i\bar{\psi}_{+}\bar{\sigma}^{\nu}\partial_{\nu}\psi_{+} -i\bar{\psi}_{-}\bar{\sigma}^{\nu}\partial_{\nu}\psi_{-}  \nonumber \\
& & + \phi_{S}\psi_{+}\psi_{-} + \phi^{\dagger}_{S}\bar{\psi}_{+}\bar{\psi}_{-} - M^{\dagger}\psi_{+}\psi_{+} -M\bar{\psi}_{+}\bar{\psi}_{+} - m^{\dagger}\psi_{-}\psi_{-} -m\bar{\psi}_{-}\bar{\psi}_{-}.
\end{eqnarray}
At the case $|M|\ne|m|$, the Lagrangian has a skew structure in the right-left symmetry space,
and thus the contributions of the right and left handed particles in quantum corrections of a loop expansion will have differences. 
The generating functional is given as
\begin{eqnarray}
Z &=& \int {\cal D}\psi_{+}{\cal D}\bar{\psi}_{+}{\cal D}\psi_{-}{\cal D}\bar{\psi}_{-}{\cal D}\phi_{+}{\cal D}\phi^{\dagger}_{+}{\cal D}\phi_{-}{\cal D}\phi^{\dagger}_{-}{\cal D}\phi_{S}{\cal D}\phi^{\dagger}_{S}\exp \Bigl[ i\int d^{4}x ({\cal L}+({\rm source}))\Bigr]   \nonumber \\
&=& \int {\cal D}\psi_{+}{\cal D}\bar{\psi}_{+}{\cal D}\psi_{-}{\cal D}\bar{\psi}_{-}{\cal D}\phi_{+}{\cal D}\phi^{\dagger}_{+}{\cal D}\phi_{-}{\cal D}\phi^{\dagger}_{-}{\cal D}\phi_{S}{\cal D}\phi^{\dagger}_{S}  \nonumber \\
& & \times \exp \Bigl[ i\int d^{4}x \Bigl( -\frac{|\phi_{S}|^{2}}{G} + \Pi^{\dagger}\Omega_{B}\Pi + \frac{1}{2}\overline{\Psi}\Omega_{F}\Psi +({\rm source}) \Bigr)\Bigr],
\end{eqnarray}
where, the matrices $\Omega_{B}$ and $\Omega_{F}$ are given as follows:
\begin{eqnarray}
\Omega_{B} &\equiv& \left(
\begin{array}{cc}
k^{2}-|\phi_{S}|^{2}-4|M|^{2}-\Delta^{2} & 2(M\phi_{S}+m^{\dagger}\phi^{\dagger}_{S})   \\
2(M^{\dagger}\phi^{\dagger}_{S}+m\phi_{S}) & k^{2}-|\phi_{S}|^{2}-4|m|^{2}-\Delta^{2} 
\end{array}
\right),  \\
\Omega_{F} &\equiv& \left(
\begin{array}{cc}
i\partfey -M^{\dagger}\frac{1+i\gamma^{5}}{2} -M\frac{1-i\gamma^{5}}{2} & -\phi_{S}\frac{1+i\gamma^{5}}{2} -\phi^{\dagger}_{S}\frac{1-i\gamma^{5}}{2}  \\
-\phi_{S}\frac{1+i\gamma^{5}}{2} -\phi^{\dagger}_{S}\frac{1-i\gamma^{5}}{2}  & i\partfey -m^{\dagger}\frac{1+i\gamma^{5}}{2} -m\frac{1-i\gamma^{5}}{2} 
\end{array}
\right),
\end{eqnarray}
( $\gamma^{5}=\gamma^{0}\gamma^{1}\gamma^{2}\gamma^{3}$ ) while the definitions of several fields in our Lagrangian are
\begin{eqnarray}
& & \Psi \equiv ( \psi_{MR}, \psi_{ML} )^{T}, \quad \psi_{MR} = (\psi_{+},\bar{\psi}_{+})^{T}, \quad \psi_{ML} = (\psi_{-},\bar{\psi}_{-})^{T}, \nonumber \\
& & \overline{\Psi} \equiv ( \overline{\psi_{MR}}, \overline{\psi_{ML}} ), \quad \overline{\psi_{MR}} = (-\psi_{+},-\bar{\psi}_{+}), \quad \overline{\psi_{ML}} = (-\psi_{-},-\bar{\psi}_{-}), \quad \Pi \equiv (\phi_{+},\phi_{-})^{T}.
\end{eqnarray}
Here, $\psi_{MR}$ and $\psi_{ML}$ are right- and left- handed Majorana fields, respectively. 
$T$ denotes transposition. $\Psi$ can be called as an eight-component Nambu-notation-field~[13] defined in terms of the Majorana fields.
( Of course, it is equivalent to the formalism of eight-component Nambu-notation given by Dirac fields which was used in relativistic theory of superconductivity~[14,15]. )
Because we have three complex mass parameters 
\begin{eqnarray}
\phi_{S} = |\phi_{S}|e^{i\theta_{S}}, \quad M = |M|e^{i\theta_{M}}, \quad m = |m|e^{i\theta_{m}}, 
\end{eqnarray}
we can absorb only two of these phases $\theta_{S}$, $\theta_{M}$ and $\theta_{m}$
by a redefinition of fields $\psi_{+}$ and $\psi_{-}$, and one phase degree of freedom remains.

\vspace{5mm}

We employ the steepest descent approximation for the integration of the collective fields $\phi_{S}$ and $\phi^{\dagger}_{S}$.
Therefore, we obtain the effective action in the following form:
\begin{eqnarray}
\Gamma_{eff} &=& \int d^{4}x\Bigl( -\frac{|\phi_{S}|^{2}}{G}\Bigr) + 2i\ln{\rm Det}\Omega_{B}-i\ln{\rm Det}\Omega_{F}.
\end{eqnarray}
The effective potential will be obtained after the diagonalizations of the matrices $\Omega_{B}$ and $\Omega_{F}$ as
\begin{eqnarray}
V_{eff} &=& \frac{|\phi_{S}|^{2}}{G} -2i\int \frac{d^{4}k}{(2\pi)^{4}} \ln\frac{(k_{0}-E^{B}_{+}(\bmk))(k_{0}+E^{B}_{+}(\bmk))(k_{0}-E^{B}_{-}(\bmk))(k_{0}+E^{B}_{-}(\bmk))}{(k_{0}-E^{F}_{+}(\bmk))(k_{0}+E^{F}_{+}(\bmk))(k_{0}-E^{F}_{-}(\bmk))(k_{0}+E^{F}_{-}(\bmk))},
\end{eqnarray}
( the diagonalization of $\Omega_{F}$ has been done by the same method given in Ref.~[14] )
where, the "quasiparticle" excitation energy spectra become
\begin{eqnarray}
E^{B}_{\pm}(\bmk) &=& \sqrt{ \bmk^{2} + |\phi_{S}|^{2} + 2|M|^{2} + 2|m|^{2} + \Delta^{2}\mp 2 \sqrt{ (|M|^{2}-|m|^{2})^{2} + |\phi_{S}|^{2}(|M|^{2}+|m|^{2}+2|M||m|\cos\Theta) } },  \\
E^{F}_{\pm}(\bmk) &=& \sqrt{ \bmk^{2} + |\phi_{S}|^{2} + 2|M|^{2} + 2|m|^{2}\mp 2 \sqrt{ (|M|^{2}-|m|^{2})^{2} + |\phi_{S}|^{2}(|M|^{2}+|m|^{2}+2|M||m|\cos\Theta) } }, \\
\Theta &\equiv& 2\theta_{S} + \theta_{M} + \theta_{m}.
\end{eqnarray}
The masses appear in these spectra show complicated structures, though the Lorentz symmetry is still kept in our theory.
( These spectra have some similarities with the mass spectra of top/bottom-stop/sbottom appear in the Minimal Supersymmetric Standard Model (MSSM)~[16-18]. )
We show several limiting cases of these spectra. 
Under $|M|=|m|$, these spectra become like
\begin{eqnarray}
E^{B}_{\pm}(\bmk) &=& \sqrt{\bmk^{2}+(|\phi_{S}|\mp 2|M|)^{2}+\Delta^{2}},   \\
E^{F}_{\pm}(\bmk) &=& \sqrt{\bmk^{2}+(|\phi_{S}|\mp 2|M|)^{2}}  
\end{eqnarray}
at $\Theta=0$. 
On the other hand, our $E^{B}_{\pm}$ and $E^{F}_{\pm}$ become 
\begin{eqnarray}
E^{B}_{\pm}(\bmk) &=& \sqrt{\bmk^{2}+|\phi_{S}|^{2}+4|M|^{2}+\Delta^{2}},   \\
E^{F}_{\pm}(\bmk) &=& \sqrt{\bmk^{2}+|\phi_{S}|^{2}+4|M|^{2}}  
\end{eqnarray}
at $\Theta=\pi$ and they coincide with the dispersion relations of the zero-chemical-potential case in the SUSY BCS superconductivity~[10,11].
In fact, the BCS-type superconductivity is a dynamical generation of a specific choice of the right- and left- handed Majorana masses. 
At $|M|\ne 0$, $|\phi_{S}|\ne 0$, $|m|=0$,
\begin{eqnarray}
E^{B}_{\pm}(\bmk) &=& \sqrt{\bmk^{2}+(|M|\mp\sqrt{|M|^{2}+|\phi_{S}|^{2}})^{2}+\Delta^{2}},     \\
E^{F}_{\pm}(\bmk) &=& \sqrt{\bmk^{2}+(|M|\mp\sqrt{|M|^{2}+|\phi_{S}|^{2}})^{2}},  
\end{eqnarray}
and thus, in the case $|M|\gg|\phi_{S}|>0$ with $|m|=0$ ( the seesaw condition ), the spectra become
\begin{eqnarray}
E^{B}_{+}(\bmk) &=& \sqrt{\bmk^{2}+\frac{|\phi_{S}|^{4}}{4|M|^{2}}+\Delta^{2}},  \\
E^{B}_{-}(\bmk) &=& \sqrt{\bmk^{2}+2|\phi_{S}|^{2}+4|M|^2+\Delta^{2}},  \\
E^{F}_{+}(\bmk) &=& \sqrt{\bmk^{2}+\frac{|\phi_{S}|^{4}}{4|M|^{2}}},  \\
E^{F}_{-}(\bmk) &=& \sqrt{\bmk^{2}+2|\phi_{S}|^{2}+4|M|^2}.
\end{eqnarray}
Hence, the seesaw mechanism suppresses the mass of the branch of $E^{F}_{+}(\bmk)$ just the same with the ordinary seesaw theory~[1-3].
In our theory, the seesaw mechanism occurs also in the scalar sector.
The choice $|m|=0$ reduces the number of mass parameters in our theory, and thus the phase degree $\Theta$ disappears from our theory.

\vspace{5mm}

After performing the Wick rotation in the momentum integration $\int d^{4}k$ of $V_{eff}$ and introducing the covariant four-momentum cutoff,
the effective potential is found to be
\begin{eqnarray}
V_{eff} &=& \frac{|\phi_{S}|^{2}}{G} + \frac{1}{16\pi^{2}}\Bigg[ 2\Lambda^{2}\Delta^{2} + \Lambda^{4}\ln\frac{(1+\alpha_{+}/\Lambda^{2})(1+\alpha_{-}/\Lambda^{2})}{(1+\beta_{+}/\Lambda^{2})(1+\beta_{-}/\Lambda^{2})}    \nonumber \\
& & -\alpha^{2}_{+}\ln(1+\Lambda^{2}/\alpha_{+}) -\alpha^{2}_{-}\ln(1+\Lambda^{2}/\alpha_{-}) 
+\beta^{2}_{+}\ln(1+\Lambda^{2}/\beta_{+}) + \beta^{2}_{-}\ln(1+\Lambda^{2}/\beta_{-}) \Bigg],   \nonumber \\
\alpha_{\pm} &\equiv& (E^{B}_{\pm}(\bmk=0))^{2}, \quad \beta_{\pm} \equiv (E^{F}_{\pm}(\bmk=0))^{2}.
\end{eqnarray}
Here, $\Lambda$ denotes the four-momentum cutoff.
$\alpha_{\pm}$ and $\beta_{\pm}$ correspond to the squares of the masses appear in $E^{B}_{\pm}$, $E^{F}_{\pm}$.
At $\Lambda^{2}\gg \alpha_{\pm},\beta_{\pm}$, one obtains
\begin{eqnarray}
V_{eff} &=& \frac{|\phi_{S}|^{2}}{G} + \frac{1}{16\pi^{2}}\Bigg[ 4\Lambda^{2}\Delta^{2}
-\alpha^{2}_{+}\ln\frac{\Lambda^{2}}{\alpha_{+}} -\alpha^{2}_{-}\ln\frac{\Lambda^{2}}{\alpha_{-}}
+\beta^{2}_{+}\ln\frac{\Lambda^{2}}{\beta_{+}} +\beta^{2}_{-}\ln\frac{\Lambda^{2}}{\beta_{-}}   \Bigg].
\end{eqnarray}
We also examine our $V_{eff}$ in the three-dimensional non-covariant cutoff scheme:
\begin{eqnarray}
V_{eff} &=& \frac{|\phi_{S}|^{2}}{G} + \frac{1}{8\pi^{2}}\Bigg[ 
  2\Lambda(\Lambda^{2}+\alpha_{+})^{3/2}-\Lambda\alpha_{+}\sqrt{\Lambda^{2}+\alpha_{+}}  
+ 2\Lambda(\Lambda^{2}+\alpha_{-})^{3/2}-\Lambda\alpha_{-}\sqrt{\Lambda^{2}+\alpha_{-}} \nonumber \\
& & 
- 2\Lambda(\Lambda^{2}+\beta_{+})^{3/2}+\Lambda\beta_{+}\sqrt{\Lambda^{2}+\beta_{+}}
- 2\Lambda(\Lambda^{2}+\beta_{-})^{3/2}+\Lambda\beta_{-}\sqrt{\Lambda^{2}+\beta_{-}}    \nonumber \\
& & 
-\alpha^{2}_{+}\ln\frac{\sqrt{\Lambda^{2}+\alpha_{+}}+\Lambda}{\sqrt{\alpha_{+}}}
-\alpha^{2}_{-}\ln\frac{\sqrt{\Lambda^{2}+\alpha_{-}}+\Lambda}{\sqrt{\alpha_{-}}}
+\beta^{2}_{+}\ln\frac{\sqrt{\Lambda^{2}+\beta_{+}}+\Lambda}{\sqrt{\beta_{+}}}
+\beta^{2}_{-}\ln\frac{\sqrt{\Lambda^{2}+\beta_{-}}+\Lambda}{\sqrt{\beta_{-}}}
\Bigg].
\end{eqnarray} 
The energy spectra of particles we have obtained are Lorentz symmetric, while the three-dimensional cutoff scheme ( usually be employed in a finite-temperature Matsubara formalism )
breaks the Lorentz symmetry explicitly.
Therefore, the self-consistent gap equation for $|\phi_{S}|$ is found from the stationary condition of the effective potential $\frac{\partial V_{eff}}{\partial |\phi_{S}|}=0$ to be
\begin{eqnarray}
\frac{\partial V_{eff}}{\partial|\phi_{S}|} &=& \frac{2|\phi_{S}|}{G}   \nonumber \\
& & - \frac{1}{8\pi^{2}}\Bigg[  
\frac{\partial\alpha_{+}}{\partial|\phi_{S}|}\Bigl\{ \alpha_{+}\ln\Bigl(1+\frac{\Lambda^{2}}{\alpha_{+}} \Bigr) - \beta_{+}\ln\Bigl(1+\frac{\Lambda^{2}}{\beta_{+}} \Bigr)    \Bigr\} 
+ \frac{\partial\alpha_{-}}{\partial|\phi_{S}|}\Bigl\{ \alpha_{-}\ln\Bigl(1+\frac{\Lambda^{2}}{\alpha_{-}} \Bigr) - \beta_{-}\ln\Bigl(1+\frac{\Lambda^{2}}{\beta_{-}} \Bigr)    \Bigr\}
\Bigg]
\end{eqnarray}
in the covariant cutoff scheme, while
\begin{eqnarray}
\frac{\partial V_{eff}}{\partial|\phi_{S}|} &=& \frac{2|\phi_{S}|}{G} +\frac{1}{4\pi^{2}}\Bigg[ \frac{\partial\alpha_{+}}{\partial|\phi_{S}|}\Bigl\{ 
\Lambda\sqrt{\Lambda^{2}+\alpha_{+}} - \Lambda\sqrt{\Lambda^{2}+\beta_{+}}
-\alpha_{+}\ln\frac{\sqrt{\Lambda^{2}+\alpha_{+}}+\Lambda}{\sqrt{\alpha_{+}}}
+\beta_{+}\ln\frac{\sqrt{\Lambda^{2}+\beta_{+}}+\Lambda}{\sqrt{\beta_{+}}}
\Bigr\} \nonumber \\
& & + \frac{\partial\alpha_{-}}{\partial|\phi_{S}|}
\Bigl\{ 
\Lambda\sqrt{\Lambda^{2}+\alpha_{-}} - \Lambda\sqrt{\Lambda^{2}+\beta_{-}} 
-\alpha_{-}\ln\frac{\sqrt{\Lambda^{2}+\alpha_{-}}+\Lambda}{\sqrt{\alpha_{-}}}
+\beta_{-}\ln\frac{\sqrt{\Lambda^{2}+\beta_{-}}+\Lambda}{\sqrt{\beta_{-}}}
\Bigr\} \Bigg]
\end{eqnarray}
in the three-dimensional cutoff regularization. 
Here, the derivatives appear in the gap equations become
\begin{eqnarray}
\frac{\partial\alpha_{\pm}}{\partial|\phi_{S}|} &=& \frac{\partial\beta_{\pm}}{\partial|\phi_{S}|} = 2|\phi_{S}|\Bigl( 1\mp\frac{|M|^{2}+|m|^{2}+2|M||m|\cos\Theta}
{\sqrt{(|M|^{2}-|m|^{2})^{2}+|\phi_{S}|^{2}(|M|^{2}+|m|^{2}+2|M||m|\cos\Theta)}} \Bigr).
\end{eqnarray}

\vspace{5mm}

Hereafter, we examine the case $|m|=0$ and thus the phase $\Theta$ disappears from $V_{eff}$ and the gap equations.
We used the numerical package ${\it Mathematica}$ ver. 6 for our calculation.
At the case $|m|=0$, the determination equation for the critical coupling in the covariant cutoff scheme is obtained as follows:
\begin{eqnarray}
G^{cr} &=& \frac{4\pi^{2}}{ (\Delta^{2}+4|M|^{2})\ln\Bigl(1+\frac{\Lambda^{2}}{\Delta^{2}+4|M|^{2}} \Bigr) - 4|M|^{2}\ln\Bigl(1+\frac{\Lambda^{2}}{4|M|^{2}} \Bigr) }.
\end{eqnarray}
From this results, we obtain $\lim_{|M|\to 0}G^{cr} = 4\pi^{2}/\{\Delta^{2}\ln(1+\frac{\Lambda^{2}}{\Delta^{2}})\}$, 
coinsides with that of the ordinary SNJL model~[8]. 
Figure 1 shows the critical coupling $G^{cr}$ as a function of cutoff $\Lambda$ under the unit $\Delta=1$.
$G^{cr}$ shows the well-known dependence on $\Lambda$, and becomes small at a large $\Lambda$.
$G^{cr}$ depends on $|M|$ quite sensitively, 
and when $|M|/\Delta$ becomes large, the critical coupling also becomes large 
and the dynamical generation of a finite VEV of $|\phi_{S}|$ will be suppressed. 
This stems from the fact that the denominator of Eq. (33) approaches to zero under the limit $|M|/\Delta\to +\infty$.
The critical coupling in the three-dimensional cutoff scheme is found to be
\begin{eqnarray}
G^{cr} &=& 2\pi^{2}\Bigg[ \Lambda\sqrt{\Lambda^{2}+4|M|^{2}} - \Lambda\sqrt{\Lambda^{2}+4|M|^{2}+\Delta^{2}}  \nonumber \\
& & \qquad \qquad + (4|M|^{2}+\Delta^{2})\ln\frac{\sqrt{\Lambda^{2}+4|M|^{2}+\Delta^{2}}+\Lambda}{\sqrt{4|M|^{2}+\Delta^{2}}}
-  4|M|^{2}\ln\frac{\sqrt{\Lambda^{2}+4|M|^{2}}+\Lambda}{2|M|} \Bigg]^{-1}.
\end{eqnarray}
This expression of $G^{cr}$ of the three-dimensional non-covariant cutoff with $|M|=0$ coinsides with the results of Refs.~[10,11] of the zero-density case.
Figure 2 shows $G^{cr}$ of the non-covariant cutoff scheme as a function of $\Lambda$.
The result of Fig. 2 is qualitatively the same with that of Fig. 1.   
In the covariant cutoff scheme, 
the second derivative of $V_{eff}$ with respect to $|\phi_{S}|$ at the origin will be given by $G^{cr}$ of Eq. (33) as follows:
\begin{eqnarray}
\frac{\partial^{2}V_{eff}}{\partial |\phi_{S}|^{2}}\Bigg|_{|\phi_{S}|=0} &=& 2\Bigl( \frac{1}{G} - \frac{1}{G^{cr}} \Bigr).
\end{eqnarray}

\vspace{5mm}

The numerical results of $V_{eff}$ of Eq. (27) by the covariant cutoff scheme are shown in Figs. 3, 4, 5 and 6, 
while that of Eq. (29) of the non-covariant cutoff scheme are given in Figs. 7, 8 and 9. 
All of our numerical calculations of $V_{eff}$ are performed under a fixed SUSY breaking scale $\Delta$: 
$\Lambda/\Delta$ is fixed to 100, and we use the energy unit $\Delta =1$.
The global minimum of $V_{eff}$ is quite sensitive to both $G$ and $|M|$, 
and thus we regard them as the relevant parameters to control a numerical value of self-consistent mass $|\phi_{S}|$.
A too large $G$ gives an unphysical solution like $|\phi_{S}|>\Lambda$.
Hence we will choose a numerical value of $G$ close to $G^{cr}$.
In our numerical calculations, we find there are cases of 
first and second order phase transitions they take place under the variation with respect to $G$. 
In the case of first order transition shown in Figs. 3, 4, 5, 7 and 8, 
they are depicted by several numerical examples of $G$, 
the global minimum of $V_{eff}$ usually locates at $|\phi_{S}|\sim{\cal O}(|M|)$ and such a $|\phi_{S}|$
is too large for satisfying the seesaw condition $0<|\phi_{S}|\ll|M|$ in our model. 
In this case, when we take a small value of $G$ to try to obtain a $|\phi_{S}|\ll|M|$,
$V_{eff}$ suddenly changes the global minimum from $|\phi_{S}|\ne 0$ into its origin $|\phi_{S}|=0$.
Hence it is difficult to obtain a small and finite VEV of $|\phi_{S}|$ which satisfies the seesaw condition. 
When $|M|$ is large, $G^{cr}$ also becomes large, 
and the energy scale of the tree level part $|\phi_{S}|^{2}/G$ becomes large for obtaining a finite VEV of $|\phi_{S}|$.
See the differences of energy scales of Figs. 3, 4, 5 and 6. 
The reason of $|\phi_{S}|\sim{\cal O}(|M|)$ stems from this fact under the first-order transition.

\vspace{5mm}

On the other hand, in the case of second order transitions given in Figs. 6 and 9, 
a global minimum of $V_{eff}$ can take an arbitrary small value of $|\phi_{S}|$ compared with the right-handed Majorana mass $|M|$ 
by choosing an appropriate numerical value of $G$, and thus the seesaw condition $0<|\phi_{S}|\ll|M|$ can be satisfied.
In this case, the theory approaches to a critical point with a continuously vanishing $|\phi_{S}|$.

\vspace{5mm}

Our model becomes the ordinary SNJL model which has no Majorana mass term, at the limit $|M|\to 0$ with fixed $\Lambda$ and $\Delta$, 
while the non-SUSY NJL model will be obtained by $\Delta/\Lambda\to\infty$.
The ordinary SNJL shows second-order transition under the variation of $G$.
Therefore, there is a boundary of the first and the second order phase transitions in the parameter space of the ratio $|M|/\Delta$.
For example, we found in our numerical calculation,
the boundary between the second and the first order transitions locates at the region $0.5<|M|/\Delta<0.55$ under the parameter choice
$|m|=0$, $\Delta=1$, and $\Lambda=100$ in the covariant cutoff scheme:
In this case, the phase transition becomes second-order at $|M|/\Delta\le0.5$, while it becomes first order at $|M|/\Delta\ge0.55$.
( Similarly, we found the boundary in the region $0.45<|M|/\Delta<0.55$ under the choice
$|m|=0$, $\Delta=1$, and $\Lambda=100$ in the case of non-covariant cutoff scheme. )
The boundary could be understood as 
\begin{eqnarray}
2|M|/\Delta\sim 1.
\end{eqnarray}
Therefore, we obtain the conclusion that the right-handed Majorana mass $|M|$ should 
take a value much smaller than the SUSY breaking mass scale $\Delta$ ( in the case $\Lambda\gg\Delta$ ), 
which is considered to have a TeV energy scale mass in modern particle phenomenology,
to satisfy the seesaw condition $|M|\gg|\phi_{S}|>0$ in our model (1).
In the case of second-order transition, we can choose $G$ close to $G^{cr}$ ( a fine-tuning procedure~[5] ),
while in the first-order case it is impossible to do the procedure to obtain a finite VEV of $|\phi_{S}|$ which satisfies 
the seesaw condition. 
At the origion of the effective potential, the second derivative $\frac{\partial^{2}V_{eff}}{\partial |\phi_{S}|^{2}}$ of (35) 
is always negative under the second-order transition, while it will take positive values under the first-order transition.
This means $G > G^{cr}$ is always satisfied for obtaining a finite VEV of $|\phi_{S}|$ in the case of second-order transition, 
while $V_{eff}$ can have a finite VEV $|\phi_{S}|\ne 0$ even if $G < G^{cr}$ under the first-order transition.
( Therefore, $G^{cr}$ lost the precise meaning under the first-order transition case. ) 
These results of $V_{eff}$ we have obtained were also confirmed in our numerical evaluation of the gap equations (30) and (31),
and also by the behavior of second derivative $\frac{\partial^{2}V_{eff}}{\partial |\phi_{S}|^{2}}$ at the origin $|\phi_{S}|=0$ of the cases of Figs. 3-9. 
The author also examined the $\Lambda$-dependence of $V_{eff}$, by choosing $\Lambda/\Delta=500$.
Still, the case $|M|/\Delta=0.1$ shows the second-order transition, 
while the case $|M|/\Delta=1$ gives the first-order transition.
Hence the order of transition is determined by the relation between $|M|$ and $\Delta$ under $|M|,\Delta\ll\Lambda$,
as indicated in (36).
Figure 10 shows the first-order transition of the example $\Delta=1$, $|M|=1000$ and $\Lambda=1000000$.

\vspace{5mm}

Our conclusions are summarized as follows:
\begin{itemize}
\item We have shown that the dynamical Dirac mass with the seesaw mechanism of neutrino can be described by our simplified model.
\item When $|M| > \Delta$ or $|M|\sim \Delta$ with $\Lambda\gg\Delta$, a self-consistent solution of $|\phi_{S}|$ will take a numerical value of the order ${\cal O}(|M|)$.
In this case, it is difficult to satisfy the seesaw condition $0<|\phi_{S}|\ll|M|$. 
\item When $|M|$ is much smaller than $\Delta$ with $\Lambda\gg\Delta$, 
we can obtain an arbitrarily small $|\phi_{S}|$ by variation with respect to $G$.
In this case, it is possible to satisfy the seesaw condition.
\end{itemize}

\vspace{5mm}

The final comment. Recently, the Nambu$-$Jona-Lasinio and Gross-Neveu models have been derived from string theory with intersecting D-branes~[19-21].
It is intersting for us whether the Majorana mass parameters we have considered here can be derived from such an attempt, and they ultimately have their origins in 
string theory, or not. In another point of view, effective actions obtained from compactifications of superstring theory has a common feature, 
no matter what method of compactification one uses~[22,23,24].
While, the MSSM can be regarded as the low-energy effective theory of the top-condensation model. 
Hence, our ultimate goal is summarized in the following scheme ( see, also~[25] ):
\begin{eqnarray}
{\rm Superstrings} \to {\rm SNJL} + {\rm Majorana} \to {\rm MSSM}.
\end{eqnarray}

\begin{figure}

\caption{The critical coupling constant $G^{cr}$ of the covariant cutoff scheme as a function of cutoff $\Lambda$.
We choose $\Delta=1$.}

\caption{The critical coupling constant $G^{cr}$ of the non-covariant cutoff scheme as a function of cutoff $\Lambda$.
We choose $\Delta=1$.}

\caption{The effective potential $V_{eff}$ ( obtained by the covariant cutoff scheme ) of the first-order phase transition as a function of 
the dynamical Dirac mass parameter $|\phi_{S}|$.
We set the parameters as $\Delta=1$, $|M|=10$, $\Lambda=100$, $|m|=0$.
In this case, we obtain $G^{cr}\Delta^{2}=17.1989$ from Eq. (33). 
The second derivative ( i.e., the curvature ) of $V_{eff}$ at the origin is negative at the case $G\Delta^{2}=17.2$, 
while it is positive at the cases $G\Delta^{2}=16.1, 16.0, 15.9$.}

\caption{The effective potential ( obtained by the covariant cutoff scheme ) of the first-order phase transition as a function of $|\phi_{S}|$.
The parameters $\Delta$, $|M|$, $\Lambda$ and $|m|$ are set as the same with Fig.3.
The positiveness of curvature of $V_{eff}$ at the origin is clear under the energy scale of this figure.}

\caption{The effective potential ( obtained by the covariant cutoff scheme ) of the first-order phase transition as a function of $|\phi_{S}|$.
We set the parameters as $\Delta=1$, $|M|=1$, $\Lambda=100$, $|m|=0$.}

\caption{The effective potential ( obtained by the covariant cutoff scheme ) of the second-order phase transition as a function of $|\phi_{S}|$.
We set the parameters as $\Delta=1$, $|M|=0.1$, $\Lambda=100$, $|m|=0$.}

\caption{The effective potential ( obtained by the non-covariant cutoff scheme ) of the first-order phase transition as a function of $|\phi_{S}|$.
We set the parameters as $\Delta=1$, $|M|=10$, $\Lambda=100$, $|m|=0$.}

\caption{The effective potential ( obtained by the non-covariant cutoff scheme ) of the first-order phase transition as a function of $|\phi_{S}|$.
We set the parameters as $\Delta=1$, $|M|=1$, $\Lambda=100$, $|m|=0$.}

\caption{The effective potential ( obtained by the non-covariant cutoff scheme ) of the second-order phase transition as a function of $|\phi_{S}|$.
We set the parameters as $\Delta=1$, $|M|=0.1$, $\Lambda=100$, $|m|=0$.}

\caption{The effective potential ( obtained by the covariant cutoff scheme ) of the first-order phase transition as a function of $|\phi_{S}|$.
We set the parameters as $\Delta=1$, $|M|=1000$, $\Lambda=1000000$ and $|m|=0$.}

\end{figure}

\end{document}